*Article*

# Estimating the Energy State of Liquids

**Lianwen Wang**

School of Physical Science and Technology, Lanzhou University, Lanzhou 730000, China; E-Mail: lwwang@lzu.edu.cn; Tel.: +86-931-8915-597

**Abstract:** In contrast to the gaseous and the solid states, the liquid state does not have a simple model that could be developed into a quantitative theory. A central issue in the understanding of liquids is to estimate the energy state of liquids. Here, on the basis of our recent studies on crystal melting, we show that the energy sate of liquids may be reasonably approximated by the energy and volume of a vacancy. Consequently, estimation of the liquid state energy is significantly simplified comparing with previous methods that inevitably invoke many-body interactions. Accordingly, a possible equation for the state for liquids is proposed. On this basis, it seems that a simple model for liquids is in sight.

## 1. Introduction

Liquids are the mother phase of most materials, from metallic, organic, to inorganic, from crystalline to amorphous. In this sense, liquids are at the center of modern Materials Science. However, until now, in contrast to the gaseous and the solid states, the liquid state does not have a simple model that could be developed into a quantitative theory [1]. A critical issue is to know the energy state of liquids because it is only after the energy state of matter is exactly known that one can predict the other properties. For gases, atomic interactions are in principle negligible. For crystals, generally only atomic vibrations are considered. In comparison, for liquids and glasses, it is agreed that many-body interactions are to be considered; unfortunately, the many-body interaction is an unresolved problem in itself.

Over the past decade, I have been working on a possible microscopic mechanism for crystal melting and the accompanying liquid nucleation. In brief, by looking into atomic migrations in pure elements, preferably towards nearby vacancies, we found that with temperature increasing atomic migrations may become so intensive that the migration atoms will finally squash the nearby vacancy and trigger local lattice instability with a liquid nucleus produced (a process that is hereafter referred to as vacancy-squashing) [2]. After this, the resulting model for liquid nucleus was tentatively applied to elucidate some silent features of liquid-to-glass transition as was summarized in [2]. Apart from crystal melting and glass transition, in the present paper, efforts are paid on the strategies of model developing for normal liquids (*i.e.*, liquids at temperatures above the melting point of corresponding crystals), particularly, (i) shall we prefer gas- or solid-like models for liquids? and (ii) shall we approach liquids in a simple or fundamental way? Accordingly, available various liquid models are critically discussed and finally our vacancy-squashing model is argued to be an efficient, solid-like, and simple model for liquids. Content in Sections 2–4 have been presented at a conference [3] and are reused here. In addition, new supporting data are added in Section 4. On the above basis, a possible equation of state for liquids is proposed in Section 5 and is verified in Section 6.

**2. Gas- or Solid-Like Models?**

In 1873, van der Waals published, in his Ph.D. thesis "On the continuity of the gas and liquid states", his famous equation of state, which takes into account both the size of and the forces among molecules:

$$\left(P+\frac{a}{V^2}\right)(V-b)=RT \tag{1}$$

where *P*, *V* and *T* are the pressure, volume and temperature of a system containing one mole of molecules, $a/V^2$ is the internal pressure caused by intermolecular forces, (*V* − *b*) is the actual available volume, and *R* the gas constant [4,5].

As was stated in the title of van der Waals' thesis, this approach to liquid focused on the continuity of the gas and liquid states and is a gas-like liquid model. The physics embodied in this model is clear and the expression is simple. In this sense, this model can be classified as a simple approach to liquids. Nevertheless, plausible as this model is, "this equation of state is of little value in predicting critical constants because the two parameters *a* and *b* are usually obtained by fitting the equation to the critical constants themselves or to low-density *PVT* data, which are not generally available for metals" [6–8]. In addition, even if the two parameters *a* and *b* are exactly known, the usage of Equation (1) for liquids should be carefully limited to temperatures not too far away from the critical point.

N. F. Mott *et al.* found from X-ray diffraction data that "the molecules of a liquid have a certain degree of order, as in a solid, and are not distributed entirely at random, as in gas". "It has therefore become fashionable to think of a liquid rather as a disordered or broken-up solid than as a highly compressed gas". Accordingly, "a liquid is nothing more than a polycrystalline solid in which the individual crystals are so small that one cannot draw any sharp distinction between the crystals and the surfaces of misfit" [9–11]. Similar approaches were later carried out by considering various crystal defects in crystals e.g., vacancies [12], dislocations [13,14] and interstitials [15]. As an empirical exercise, such models are watertight; what the models lack is a mechanism [16]. These considerations estimated the dependence of the free energy of a crystal against the concentration of crystal defects, from which the crystal-liquid transition was somehow assigned. However, strictly speaking, what these models produced at the instance of melting was indeed "a crystal saturated with defects" rather than "a liquid". In other words, in the above models, some energy did have been estimated, but not necessarily for a liquid.

H. Eyring *et al.* [17] and J. Frenkel [18] did not pay too much attention to the crystal-liquid transition and went directly to calculations of the energy state of "a system with excess volume (fluidized vacancies or holes)". Although such a system they considered is very alike to "a liquid", these two approaches encountered the difficulty of quantifying the energy of atomic migration [17] or hole formation [18]. As a result, these two models remain valuable qualitative and explanative theories of liquids rather than quantitative and predictive ones.

The reason for the partial failure of the liquid models by Eyring *et al.* and by Frenkel is, in my view, that the two models are not concluded from (although they did have focused on [19,20]) the continuity of the crystal and liquid states; while the reason for the partial success of the gas-like model for liquids by van der Waals is that the model quantified the continuity of the gas and liquid states. Essentially, the models by Eyring *et al.* and by Frenkel, though have quite valuable insights into liquids, are not real solid-like approaches to liquids. Meanwhile, the real solid-like approaches by Mott *et al.* and by others just did not see the real liquid. As such, towards a further understanding of liquids, one will inevitably need to work on the continuity of the crystal and liquid states.

## 3. Simple or Fundamental Approaches?

Aside from the above model approaches, which are relatively simple, there was a group of fundamental approaches to the energy state of liquids by e.g., M. Born and H. S. Green [21–23] (please refer to Section 2 of Ref. [1] for a brief review), which tried to integrate all the possible forces exerted on each of the molecules [9,24]:

$$\int e^{-W(q_1, q_2 \ldots)/kT} dq_1, dq_2 \ldots \quad (2)$$

where $W(q_1, q_2\ldots)$ is the potential energy of the atoms of a system in a configuration defined by the coordinates $q_1, q_2\ldots$ and $k$ is Boltzmann's constant.

Evaluation of the above integral is, however, so complicated for any but the simplest forms of potential that various assumptions are inevitably needed for practical purposes. In this context, I shall borrow the words by W. P. Boynton and A. Bramley [25] (on the developing of an equation of state) that "to be of value an equation must describe the behavior of actual substances to some degree of approximation, and must be mathematically tractable. If it fails in the first point, it is worthless; if in the second, useless. It is more satisfying if it also has a rational basis".

The major difficulty encountered in calculating Equation (2) is the so-called many-body problem, which is unresolved in itself, see Figure 1a. Open questions are: for calculating $W(q_1, q_2\ldots)$, shall we use the pair interatomic potential, or the triplet potential? Shall we integrate the interatomic forces among the first shell, or the second shell? Or, even more? (For a detailed discussion on this issue please refer to Section 2 of Ref. [1] and references therein. To avoid possible confusions, it is noted here that the "pair" and "triplet" are not the order of a Virial equation of state [9]).

To simplify the calculations of Equation (2), J. E. Lennard-Jones [24] supposed that "each atom is confined to a cell by the forces of its neighbors, and the field surrounding it is calculated on the assumption that the neighbors are in their equilibrium positions." Accordingly, a free volume was defined as an integral over an atomic cell, with $\phi$ being the potential energy of the field within it, see Figure 1b:

$$v_f = \int e^{-\phi/kT} \mathrm{d}v \quad (3)$$

It is noted that, according to Ref. [24], Equation (3) is a certain weighted average of the volume within the atomic cell (dashed shadowed circle in Figure 1b). "Owing to the high positive values which $\phi$ takes when atoms enter each other's repulsive fields, the integral rapidly approaches zero and the integral is a small, finite fraction of the actual volume occupied by each atom" [24].

**Figure 1.** Two-dimensional diagram for an ensemble of atoms in liquids or glasses: (**a**) traditionally, the system was considered as a many-body problem; (**b**) in Lennard-Jones' approach, the potential energy of the field within an atomic cell (dashed shadowed circle) was integrated; (**c**) in our liquid model, all the forces exerted on each of the atoms is estimated by using the energy needed to produce a vacancy (dashed empty circle).

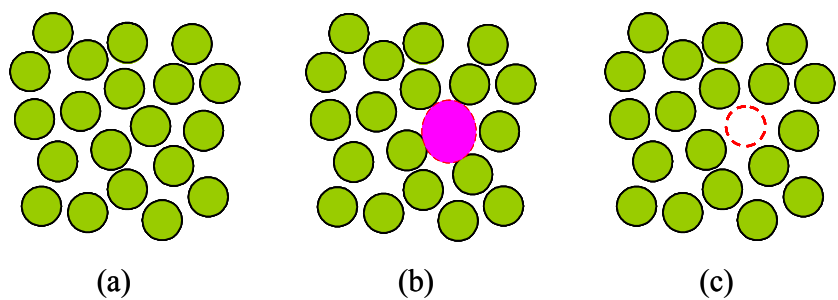

(a) (b) (c)

Comparing with Equation (2), Equation (3) is mathematically much more tractable. However, at this stage, the model by Lennard-Jones is just as qualitative as the models by Eyring *et al.* and by Frenkel because $\phi$ and hence the defined free volume cannot be quantified a priori.

The trend in simplifying the over-complicated fundamental approach to liquids is obvious. The Boynton-Bramley criterion [25] for an equation of state reflected the need for a simple at least tractable model. While Occam's razor advocates the simplest solution, Einstein's razor warns against too much simplicity. For liquids, what we need should of course be a model as simple as those for gases and crystals.

In our liquid model, the forces exerted on each of the molecules is estimated not by integrating the forces, as in Equations (2) and (3), but by using the energy needed to take a molecule away *i.e.*, to produce a vacancy, see Figure 1c. Consequently, the volume of a vacancy may in a sense correspond to the free volume in the models by Eyring *et al.*, Frenkel and Lennard-Jones. Two questions may arise instantly: (i) In using the energy to take a molecule away, why the energy of vacancy formation, not the energy of evaporation, is used? (ii) How the energy and volume of a vacancy are distributed among the molecules? These issues will be discussed in Section 4. What I want to stress here is that, by using the activation energy of vacancy formation to estimate the energy state of liquids, our liquid model avoided the integrations like Equations (2) or (3) and hence is quite simple. On the basis of above discussions, hope our liquid model were not an over-simplified one.

## 4. On the Continuity of the Crystal and Liquid States

In this section, the two questions raised in the preceding paragraph will be discussed. Eyring [26] took for granted that "to form a hole the size of a molecule in a liquid requires almost the same increase in free energy as to vaporize a molecule". This is actually not the case: to vaporize a molecule may not necessarily produce a vacancy and vice versa. To produce a vacancy merely results in an expansion of about a molecular volume in a crystal or liquid, in comparison, by vaporizing a molecule, the volume of a crystal or liquid decreases by a molecular volume with the vaporized molecule brought into the corresponding gas phase. Nowadays, for elemental metals, it is possible to directly compare the measured enthalpies for vacancy formation [27] and vaporization [28]. It turns out that the enthalpy of vacancy formation is around one third of that for vaporization. What we are interested here in Figure 1c is the formation of a vacancy rather than vaporization of a molecule and consequently the energy of vacancy formation is used for the estimation of the energy of liquids.

The second question is how the energy and volume of a vacancy are distributed among the molecules? This is directly related with the microscopic mechanism of melting, see Figure 2, *i.e.*, what the latent heat and volume change of melting are for?

**Figure 2.** A preferable approach to liquids should be a crystal-like simple model introduced on basis of the continuity of the crystal and liquid states (indicated by dashed circle) at the instance of crystal melting (indicated by arrow). For this purpose, the critical issue is to interpret microscopically how the latent heat $\Delta H_m$ and volume change $\Delta V_m$ of melting connect a crystal and corresponding liquid.

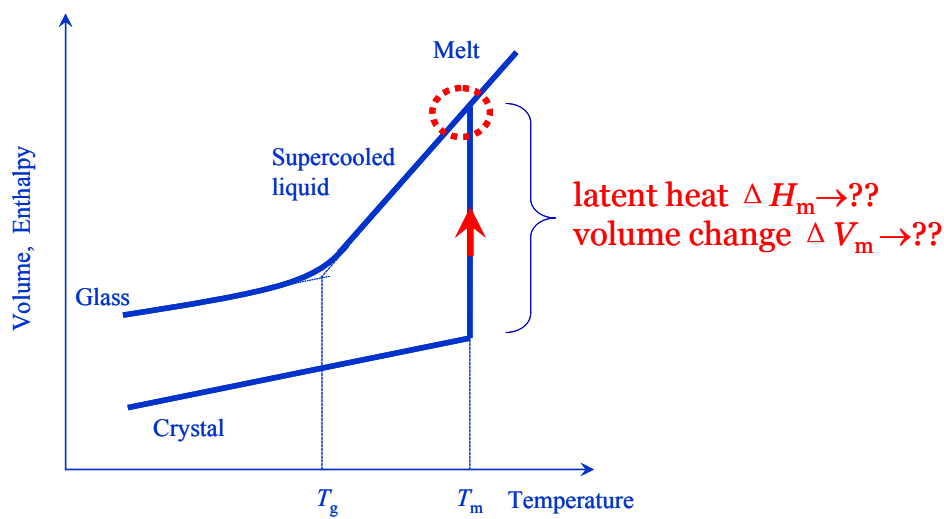

In Ref. [29] we found that the latent heat $\Delta H_m$ and volume change $\Delta V_m$ of melting corresponded to the energy and volume of a vacancy, upon squashed, stored in the system and divided by the first coordinating atoms. Specifically, $\Delta H_m = 2E_m/N$ and $\Delta V_m/V_S = V_F/(N\Omega/\eta)$, where $N$ is the number of first coordinating atoms, $E_m$ the vacancy migration energy (*i.e.*, the activation energy needed for an atom to jump into a neighboring vacancy), $V_S$ the volume of crystals before melting, $V_F$ and $\Omega$ respectively the volume of a vacancy and of an atom, and $\eta$ the packing density in the crystal e.g., 0.68 for bcc structure and 0.74 for fcc/hcp structure.

Here, to further support the conclusion drawn in Ref. [29], the data in Ref. [29] are extended to include the so-called "abnormal" elements such as Si, Ge, Sb and Bi which are thought to have "anomalously large" enthalpy changes and contract in volume during melting. See Tables 1 and 2 for the data on the latent heat and volume change during melting.

**Table 1.** The activation energy of vacancy migration $E_m$ [30], the number of the first coordinating atoms in crystals $N$ [31], the accordingly calculated latent heat of melting $2E_m/N$ [29], and the measured latent heat of melting $\Delta H_m$ [28].

|    | $N$ | $E_m$ (kJ/mol) | $2E_m/N$ (kJ/mol) | $\Delta H_m$ (kJ/mol) |
|----|-----|----------------|-------------------|------------------------|
| Ar | 12  | 10.1           | 1.7               | 1.18                   |
| Kr | 12  | 12.6           | 2.1               | 1.64                   |
| Xe | 12  | 20.6           | 3.4               | 2.27                   |
| Si | 4   | 105.0          | 52.5              | 50.2                   |
| Ge | 4   | 94.4           | 47.2              | 36.9                   |
| Bi | 5   | 48.2           | 19.3              | 11.1                   |
| Sb | 5   | 28.9           | 11.6              | 19.8                   |
| In | 12  | 29.9           | 5.0               | 3.3                    |

**Table 2.** The number of the first coordinating atoms in crystals $N$ [31], the packing density in crystals $\eta$ [31], the vacancy formation volume $V_F$ as estimated from the activation volume of self-diffusion in crystals $V^*$ [32] in terms of atomic volume $\Omega$, the calculated volume change during melting $\Delta V_m/V_S = V_F/(N\Omega/\eta)$ [29], where $V_S$ is the volume of crystals before melting, and the measured $\Delta V_m/V_S$ data [33].

|    | $N$ | $\eta$ | $V_F \approx V^*$ ($\Omega$) | Calculated $\Delta V_m/V_S$ (%) | Measured $\Delta V_m/V_S$ (%) |
|----|-----|--------|-------------------------------|----------------------------------|--------------------------------|
| Si | 4   | 0.34   | −1.42                         | −12.1                            | −9.5                           |
| Ge |     |        | −0.75                         | −6.4                             | −5.1                           |
| Bi | 5   | 0.44   | −0.36                         | −3.2                             | −3.9                           |
| Sb |     |        | −0.07                         | −0.6                             | −1.0                           |
| Li | 8   | 0.68   | 0.26                          | 2.2                              | 2.7                            |
| Na |     |        | 0.41                          | 3.5                              | 2.6                            |
| K  |     |        | 0.55                          | 4.6                              | 2.5                            |
| Rb |     |        | 0.40                          | 3.4                              | 2.3                            |
| Tl |     |        | 0.55                          | 4.7                              | 2.2                            |
| Pu |     |        | −0.34                         | −2.9                             | −2.4                           |
| Ce |     |        | −0.09                         | −0.8                             | −1.0                           |
| Eu |     |        | 0.68                          | 5.8                              | 4.8                            |
| La |     |        | 0.10                          | 0.9                              | 0.6                            |
| Er |     |        | 1.39                          | 11.8                             | 9.0                            |
| Yb |     |        | 0.59                          | 5.0                              | 4.8                            |

Figure 3 shows the comparison between the measured [28] and calculated latent heat of melting for Si, Ge, Bi, Sb, In and rare-gas solids Ar, Kr and Xe [30]. Considering measurement inaccuracies in $\Delta H_m$ and particularly in $E_m$, both of which are difficult to measure, the agreement is actually good.

**Figure 3.** Coincidence of the calculated and measured latent heat of melting for Si, Ge, Bi, Sb, In (filled squares) and Ar, Kr, Xe (filled circles) as complied in Table 1. Data reported in [29] were also shown for reference (open circles and squares).

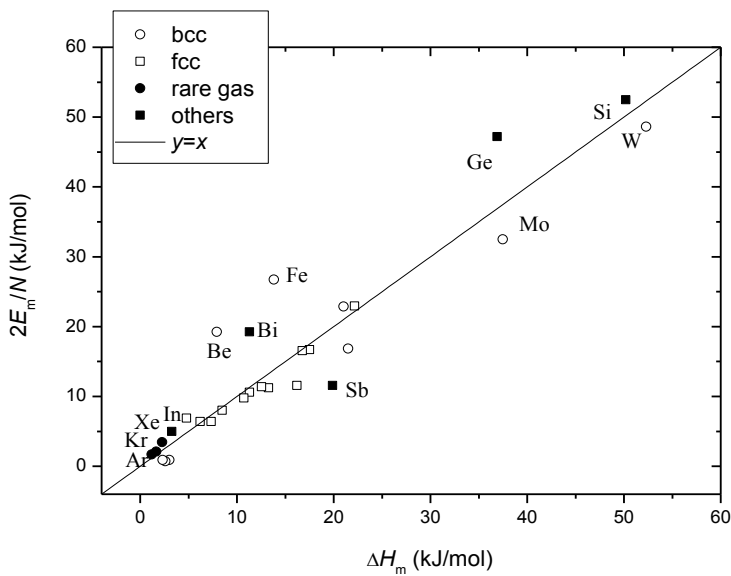

**Figure 4.** Agreement of calculated and measured volume changes on melting for Si, Ge, Bi, Sb, Pu, Ce, La, Tl, Eu, Yb, Er as compiled in Table 2. Data reported in Ref. [29] are also shown for reference (open circles).

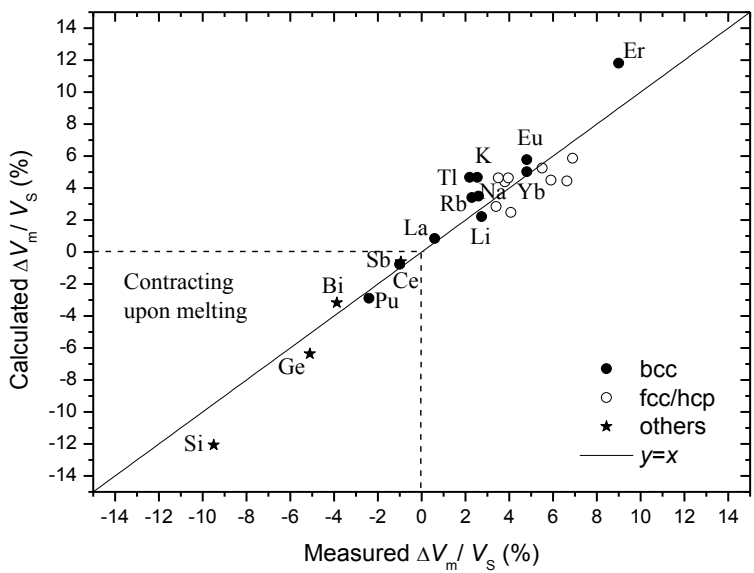

The data on the volume change on melting reported in Ref. [29] are expanded to include Si, Ge, Bi, Sb, Pu, Ce, La, Tl, Eu, Yb, Er. Here the volume of a vacancy, $V_F$, is estimated using the activation volume of self-diffusion in crystals $V^* = V_F + V_{mig}$ [32] where $V_F$ is in the order of atomic volume $\Omega$ and the migration volume $V_{mig}$ is around 0.05~0.2 $\Omega$ [34]. Possible measurement inaccuracies in $V^*$ are considered and for simplicity and for a rough estimation, we use $V_F \approx V^*$ in the present paper. The accordingly calculated and reported [33] volume changes during melting are compared in Figure 4. A good agreement is seen between the calculated and the measured volume changes on melting for elements that both expand and contract on melting, indicating the validity of the vacancy-squashing model to these elements with a variety of crystal structures.

On the basis of Ref. [29] and the supporting data added here, the question shown in Figure 2 with regard to "how the energy and volume of a vacancy are distributed among the molecules" can be to a certain extent answered: at the instant of melting when a vacancy was squashed by neighboring intensive migration atoms, the energy and volume of a vacancy are distributed among the first coordinating atoms to form a small liquid nucleus.

It is noted that this conclusion is simply from energy or thermodynamic considerations and, before any possible reconciliation with kinetic results could be revealed, seems over-simplified.

Actually, it has to be pointed out that, in view of a melting model, our work is in a quite preliminary stage, considering only the superheated melting point $T^*$, specifically, (i) what factor may determine $T^*$ and (ii) why? For (i), a possible factor was argued to be intensive atomic migrations and resulting vacancy-squashing in crystals at high temperatures [35]. Following this, a question arises instantly, namely, (ii) why vacancy-squashing should correspond to a melting transition, or in other words, what it may imply that vacancy-squashing corresponded to a melting transition. A possible explanation for this was that a liquid might have been produced upon vacancy-squashing [29]. The above are the two arguments that we made on the issue of melting. While the answer to (ii) was simply of energy considerations or of thermodynamics, the answer to (i) was, I think, somehow of kinetics.

Unfortunate is that so far our model has not been further developed to include e.g., the kinetics or statistics of liquid phase growth as was demonstrated in Ref. [36,37]. For doing so with our model, the first difficulty expected to encounter is an atomic-scale definition of the solid-liquid interface, which seems unrealistic at present.

Nevertheless, fortunate is that the simple energy considerations of liquid nucleation in our model were in accordance with available research results on liquid nucleation including Ref. [37], as were summarized in Ref. [2]. Particularly, the kinetic results in Ref. [37] showed that, during melting, the size of the critical liquid nucleus $n^*$ "for Cu is about 22 strict liquid atoms; it should be as high as 140 atoms if some quasi-liquid atoms within the interface are taken into account. The latter value is in agreement with the classical nucleation theory when the interface is considered." In addition, the estimation by our model [29] "possibly represents a lower limit of $n^*$." (see Ref. [37], page 9, left column, 1st paragraph). That is to say, the controversy over the critical nucleus was only apparent.

As such, we argue that, on the way to develop a simple model for liquids, one that as simple as those for gases and crystals, our model is laudable rather than over-simplified. In addition, it is interesting to see that the model was successfully applied to several silent features of glass-forming liquids [2] and the size effect of melting [38]. It seems that a simple model for liquids is in sight.

## 5. A Possible Equation of State for Liquids

It is only on the basis of the above discussions that I dare open up this section to search for a possible equation of state for liquids in view of our vacancy-squashing model for melting and liquid nucleation. In essence, we tried to estimate the energy state of liquids by using the energy and volume of a vacancy shared by the first shell atoms. Normally, the difference between the densities of a liquid and its solid is within 10% [33] and in this sense liquids are more like solids than gases. Consequently, properties of vacancies in a liquid are assumed to be similar to those in the corresponding crystal. The concentration of vacancies in liquids is written as:

$$C_v = \exp\left(\frac{S_v}{R} - \frac{E_v + PV_F}{RT}\right) \quad (4)$$

where $S_v$ and $E_v$ respectively are the entropy and activation energy of vacancy formation, $V_F$ the volume of a vacancy, $R$ the gas constant, $P$ the pressure and $T$ the temperature.

To make Equations (4) and (1) comparable, they are respectively reformulated as:

$$E_v + PV_F = (S_v - R \ln C_v)T \quad (5)$$

and

$$\frac{a}{V^2}(V-b) + P(V-b) = RT \quad (6)$$

As were defined by van der Waals [4], $b$ is four times the atomic/molecular volume $\Omega$, $V$ molar volume of gases, and $(V - b)$ the actual volume available for molecules. The physical meaning of $(V - b)$ is the same to the free volume defined in Equation (3) and Figure 1b and the volume of a vacancy $V_F$ in our model. Then the item $a(V - b)/V^2$, the so-called internal pressure caused by intermolecular forces times the free volume, is equivalent to the formation energy of vacancies.

Quantitatively, at the critical point $T_c$, where the critical volume $V_c \sim 3b$, the item $(V_c - b)$ is about 8 $\Omega$. For a system that have 12 first coordinating atoms in the crystal state e.g., argon, $(V_c - b) \sim 8\,\Omega$ may mean that, at $T_c$, 8 of the 12 atomic volumes are empty, *i.e.*, $C_v(T_c) \sim 8$. This is in agreement with the experimental result on liquid argon that the first coordination number reduced to about 4 at temperatures near $T_c$ [39].

Additionally, comparing Equations (5) and (6) at $T_c$, we have

$$S_v - R \ln C_v = \frac{1}{8}R \quad (7)$$

If $C_v(T_c) \sim 8$, then $S_v(T_c)$ is about $2R$. This value is comparable with the vacancy formation entropy at the melting point $S_v(T_m)$ which is about $4R$ [35].

## 6. Surface Tension in Equilibrium with the Energy of a Vacancy in Liquid Metals

A commonly used method for testing an equation of state is to compare with thermodynamic properties e.g., the temperature and pressure dependences of volume and heat capacity. For doing so, further according derivations of the proposed Equation (5) are needed on the one hand, and on the other hand, a thorough collection and evaluation of available relating experimental results is to be worked out. Some preliminary work is being done along this direction, which however may to a certain extent go out of the scope of the present manuscript on the continuity of crystal and liquid states and the estimation of the energy state of liquids.

Despite this, recently, I found a possible alternative and relatively direct way to support the arguments presented in this manuscript. If the energy state of liquids could be estimated by using the energy and volume of a vacancy, then for a liquid nucleus the surface tension should be in straightforward equilibrium with the energy of a vacancy $E_v$ (see Figure 5):

$$E_v = \frac{2\gamma_L}{r} \frac{4}{3}\pi r_v^3 \qquad (8)$$

where $\gamma_L$ is the surface tension of liquids, $r_v$ radius of the vacancy, $(2\gamma_L/r_v)$ the excess pressure over a curved surface, coinciding the so-called internal pressure item $a/V^2$ in Equation (6), and $(4\pi r_v^3/3)$ the volume of a vacancy $V_F$. At temperatures near the melting point, $r_v$ can be conveniently approximated by atomic radius $r$, and hence $V_F$ by an atomic volume $\Omega$.

**Figure 5.** Two-dimensional diagram for a snapshot of the proposed liquid nucleus, namely, a small number of atoms surrounding a vacancy with the forces exerted by the atoms (*i.e.*, the surface tension $\gamma_L$ indicated by the dashed arrows) in equilibrium with the energy of a vacancy (dashed empty circle).

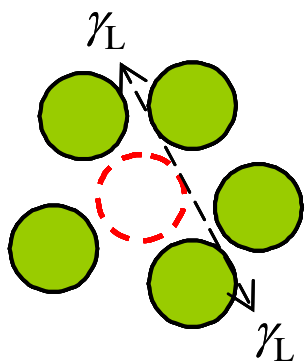

In Figure 6 the validity of Equation 8 is verified by comparing the measured vacancy formation energy $E_v$ [27] with the calculated $E_v$ by using the reported surface tension $\gamma_L$ for liquid metals [33] and atomic radii $r$ data [40]. It is interesting to see that, at least for most of the elements, the macroscopically measured surface tension is indeed closely related with the microscopic vacancy formation energy. Furthermore, several apparent exceptions are noted for which additional considerations are needed. For Sb, the reported surface tension is 367 mN/m, significantly lower than e.g., In (556 mN/m) and Sn (560 mN/m). For Mo, Ta, Nb and W that have melting points near or above 3000 K, both $\gamma_L$ and $E_v$ may inevitably suffer from measurement inaccuracies.

The correlation between $\gamma_L$ and $E_v$ may hardly be taken as merely casual. Actually, for non-metallic small molecular substances, it is an old tradition to estimate $\gamma_L$ with the internal pressure data derived from measured thermodynamic properties [41,42]. Again should I remind the reader that, by comparing Equations (5) and (6), we argued in Section 5 that the product of the so-called internal pressure $a/V^2$ caused by intermolecular forces and the free volume $(V - b)$ is equivalent to the formation energy of vacancies $E_v$. In Equation 8, the internal pressure term is estimated by using surface tension of liquids and the free volume term by using the volume of a vacancy.

**Figure 6.** Measured vacancy formation energy $E_v$ [27] in comparison with the calculated $E_v$ according to Equation (8) by using the reported surface tension $\gamma_L$ for liquid metals [33] and atomic radii $r$ data [40].

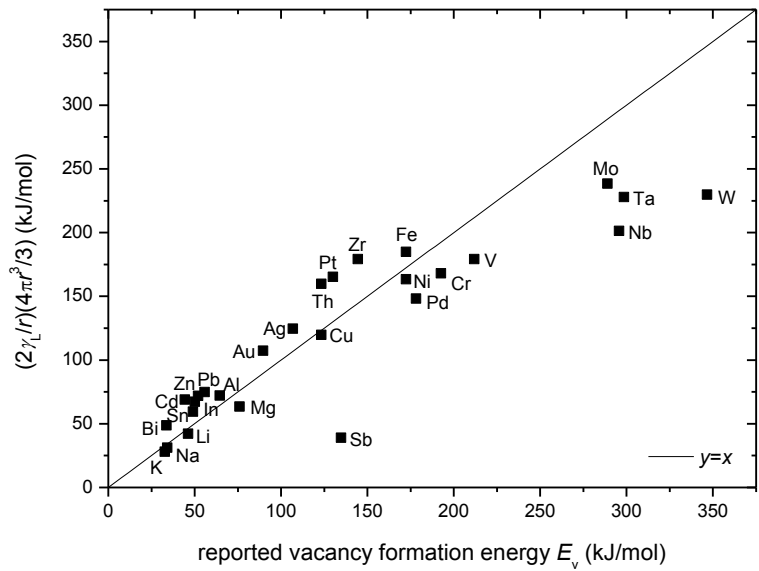

Still I may go further with Equation (8) to calculate the temperature dependence of surface tension. It is known that both surface tension $\gamma_L$ and atomic distance $d$ show significant temperature dependence; however the vacancy formation energy $E_v$ does not. After Equation 8 is rewritten as

$$\gamma_L(T) = \frac{3E_v}{8\pi} r_v(T)^{-2} \qquad (9)$$

the temperature dependence of $\gamma_L$ is expressed in the following form:

$$\frac{d\gamma_L}{\gamma_L dT} = -2 \frac{dr_v}{r_v dT} \qquad (10)$$

From the reported data on $d\gamma_L/dT$ and $\gamma_L$ [33], the temperature coefficient of surface tension $d\gamma_L/(\gamma_L dT)$ is easily obtained. The temperature coefficient of atomic distance $d$ can be derived from the reported temperature coefficient of the density $\rho$ (or volume $V$) of liquid metals [33]. In addition, it is noticed from Figure 5 that thermal expansion in the vacancy should embody all the thermal expansion of the liquid nucleus as a whole. In one dimension, $dr_v$ contains not only the thermal expansion of the vacancy in itself, but also the expansion for two neighboring atoms. That is to say, for Equation (9), we should use

$$\frac{dr_v}{r_v dT} = 3\frac{dd}{d\,dT} = \frac{dV}{V dT} = -\frac{d\rho}{\rho dT} \qquad (11)$$

Accordingly, the temperature coefficients of surface tension and density are also straight forwardly correlated by Equations (9), which is supported by experimental data [33] as are shown in Figure 7.

**Figure 7.** Comparison of measured temperature coefficient of surface tension $d\gamma_L/(\gamma_L dT)$ [33] and those calculated according to Equation (9) by using reported thermal expansion data for liquid metals [33]. Inaccuracies are large in the measured $d\gamma_L/dT$ and for most elements a value range was given. Plotted in this figure are the average of the maximum and minimum (indicated by error bars) of the given value range.

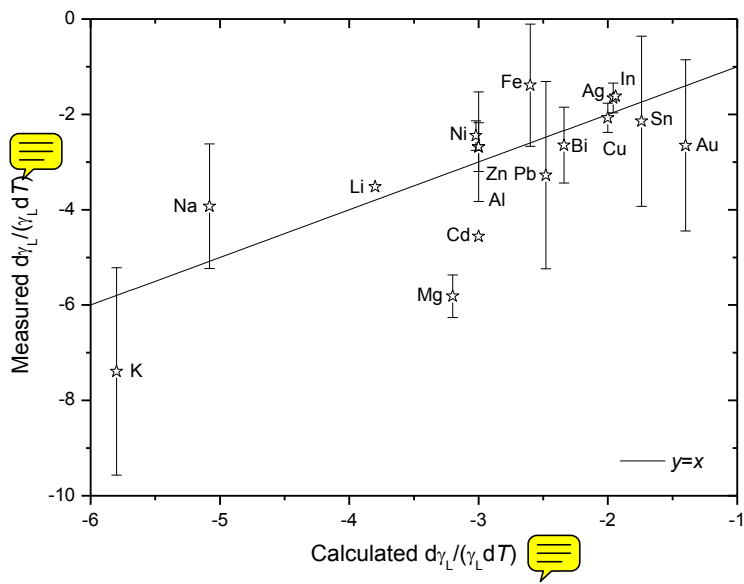

To sum up this section, in addition to the correlation between vacancy and crystal melting discussed in Section 4, from which a possible equation of state for liquids is concluded (Equation (5)), the connection of vacancy with surface tension of liquids may lend additional support to Equation (5) as well as its qualitative equivalence with the van der Waals' equation for real gases (Equation (6)).

## 7. Discussion: A Solid-Liquid Critical Point?

When discussing the continuity of crystal and liquid states, it is necessary to consider the possibility of a solid-liquid critical point. The existence of a possible solid-liquid critical point is still under debates [43].

However, there did exist detailed discussions [44,45] on the gradual evolutions of the Gibbs free energy of a model crystal with the amount of defects (considered in terms of vacancies) it contains, see Figure 8 [45]. At temperatures near the melting point, the concentration of vacancies $C_v$ in metal elements is about $10^{-3}$, which is negligible for free energy considerations, and melting occurs at the measured equilibrium melting point $T_m$. With $C_v$ increasing, the Gibbs free energy of the crystal increases and hence the melting point should decrease accordingly, which finally produces a solid-liquid critical point $T^*$ at $C_v^*$. To be noted is that, below a so-called idea glass transition temperature $T_{g0}$, increment in vacancies can only turn the crystal into an amorphous solid.

**Figure 8.** Reproduced Figure 2 of Ref. [45]. Universal melting diagram representing three nonequilibrium states: the crystal supersauterated with vacancies, the undercooled liquid and the amorphous solid. Additionally, included as a function of the respective crystal defect concentrations, $C_v$, are the (crystal-liquid) isentropic ($\Delta S^* = 0$) and isenthalpic temperatures ($\Delta H^* = 0$) which cross the melting line ($\Delta G^* = 0$) at the melting instability ($T^*$, $C_v^*$).

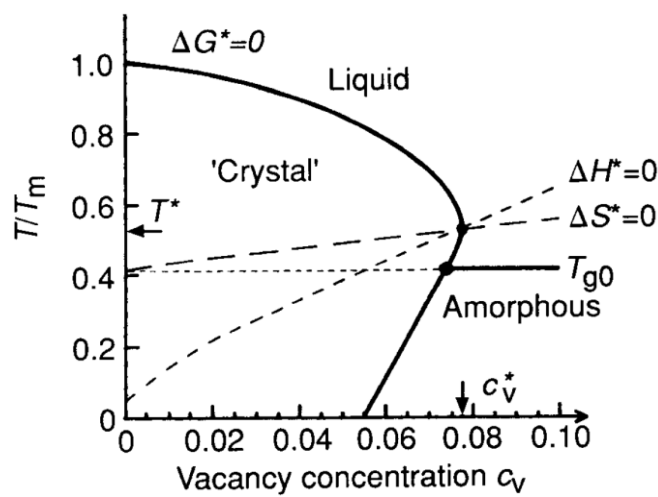

Experimentally, such a solid-liquid critical point is not easy to detect because when subjected to hydrostatic pressures most elements contract in volume, *i.e.*, the equivalent vacancy concentration $C_v$ decreases rather than increases. However, for systems that have negative vacancy formation volumes, e.g., Si, increasing the hydrostatic pressure may be taken as an increasing in $C_v$ and depression in the melting point [46] or crystal-amorphous transition [47] did have been measured under compression.

Back to the present microscopic estimation of the energy state of liquids, it will be intriguing although challenging if some microscopic details can be added to Figure 8 in addition to available macroscopic considerations.

## 8. Conclusions

On the basis of critical discussions on the strategies of liquid model developing, our crystal-like simple model for liquids is preferred, in which the energy state of liquids is estimated by employing the energy and volume of vacancies. Supporting data are provided. Finally, a possible equation of state for liquids is proposed.

## Acknowledgments

This work was supported by the National Natural Science Foundation of China (grant No. 51101077), the Scientific Research Foundation for the Returned Overseas Chinese Scholars, State Education Ministry, and the Natural Science Foundation of Gansu Province, China (grant No. 1208RJYA046).

## Conflicts of Interest

The author declares no conflict of interest.